\documentstyle[prb,aps,graphicx]{revtex}  
\def\CD{ClMePD-DMeDCNQI}        
\def\C{ClMePD}                  
\def\D{DMeDCNQI}                
\def\cm{cm$^{-1}$}               
\begin{document}

\title{A new type of neutral-ionic interface in mixed\\ stack
organic charge transfer crystals:\\ Temperature induced ionicity change
in\\ ClMePD-DMeDCNQI}

\author{Matteo Masino and Alberto Girlando}
\address{Dip. Chimica genrale ed Inorganica, Chimica Analitica e 
Chimica Fisica, Universit\`a di Parma, Parco Area delle Scienze,
43100-I Parma, Italy}

\author{Luca Farina and Aldo Brillante}
\address{Dip. di Chimica Fisica ed Inorganica, Universit\`a di
Bologna, Viale Risorgimento 4, 40136-I Bxologna, Italy}

\maketitle

\begin{abstract}

Raman and polarized infrared spectra of the mixed stack charge
transfer crystal
2-chloro-5-methyl-$p$-phenylenediamine--2,5-dimethyl-dicyanoquinonediimine
({\CD}) are reported as a function of temperature. A detailed
spectral interpretation
allows us to gain new insight into the temperature induced
neutral-ionic transition in this compound. In particular,
the crossing of the neutral-ionic borderline appears to be
quite different from that of the few known temperature induced
neutral-ionic phase transitions. First of
all, the ionicity change is continuous.
Furthermore, the onset of stack dimerization precedes,
rather than accompanies, the neutral-ionic crossing.
The (second order) phase transition is then driven
by the dimerization, but the extent of dimerization is
in turn affected by the ionicity change.

\end{abstract}


\twocolumn

\section{Introduction}
Pressure or temperature induced neutral to ionic phase transitions
(PINIT and TINIT, respectively) in mixed stack organic charge
transfer (CT) complexes were experimentally discovered about
twenty years ago.\cite{torrance1981}
The discovery, which followed earlier theoretical predictions,
\cite{mconnell1965} triggered intense experimental and theoretical
investigations. However, the experiments were slowed down by the
difficulties connected with high pressure studies, since for a long time
tetrathiafulvalene-chloranil (TTF-CA) was the only complex 
undergoing TINIT. On the other hand, theoretical interest declined
after a satisfactory and rather complete interpretatation of the 
then available experimental data was reached. \cite{VB} 

Renewed interest on NIT has been recently prompted by the
discovery that the transition can also be photoinduced,
thus shifting both the experimental and
theoretical focus to the dynamical aspects of the phase
transition and of its precursor regimes.\cite{koshihara1999}
Furthermore, it has been pointed out that since at the NIT
the ionicity jump is accompanied by a dimerization of the
mixed stack, the
transition is at least potentially of ferroelectric type.\cite{cailleau1997}
This connection puts NIT into a new 
and interesting perspective.

As mentioned above, very few compounds exhibit\linebreak TINIT in addition
to (room temperature) PINIT, thus allowing more detailed
experimental studies. Apart from \linebreak TTF-CA and its variants,
like dimethyltetrathia\-fulvalene--chloranil~\cite{aoki93}
or tetraselenofulvalene--chloranil,\cite{horiuchi1998}
only two other CT complexes have been reported to undergo TINIT,
namely tetramethylbenzidine--TCNQ,~\cite{iwasa1990} and more recently  
2-chloro-5-methyl-{\it p}-phenylenediamine
--2,5-dimethyl-dicyanoquinonediimine,
{\CD}.~\cite{aoki}

Aim of the present paper is to obtain a more detailed 
characterization of the TINIT in {\CD}, through micro-Raman 
and polarized single crystals infrared (IR) studies.
As shown below, in {\CD} the crossing of the 
neutral-ionic borderline is different from the other known cases. 
First of all, there is no appreciable jump of ionicity at the 
phase transition. Furthermore, the onset of stack dimerization appears 
well before the neutral-ionic crossing.

\section{Experimental}

{\CD} crystals have been prepared by mixing saturated solutions
in dichlorometane followed by slow evaporation of the
solvent.\cite{aoki} The IR spectra have been obtained with a
FT-IR spectrometer equipped with a microscope (Bruker IFS66 with
A590 microscope). The MCT detector employed in this arrangement
limit the low frequency side of the IR spectra to 600 {\cm}.
The Raman spectra were recorded with a Renishaw System 1000
Microscope with excitation from an Ar ion laser ($\lambda$ = 514.5 nm).
For both low (down to 80 K) and high temperature 
measurements we have employed a Linkam HFS91 cold stage.

\section{Results}

\subsection{Spectral Predictions}
{\CD} crystallizes in the triclinic system, space group P1,
with a=7.463\AA, b=7.504\AA, c=7.191\AA, $\alpha$=91.23$^{\circ}$, 
$\beta$=112.19$^{\circ}$, $\gamma$=96.91$^{\circ}$, and Z=1.~\cite{aoki} 
The crystal structure is characterized by the presence of mixed
stack columns along the $b$ axis, where the electron-donor (D)
{\C} and the electron acceptor (A) {\D} alternate along the stack.
The details of the crystal structure are not available,\cite{aoki}
but it appears that at room temperature the stack is regular, that is,
each molecule has equal distance (CT integral) with its two nearest neighbors
along the chain. Moreover, the stack structure presents disorder in the 
relative orientation of the two molecules: the 2- and 5-
substitutional sites of {\C} are considered to be occupied by
chlorine and methyl group with equal probability along the stack.

Vibrational spectroscopy is an important tool to study the ground 
state structural properties of CT compounds
and of NIT in particular, since the normal modes 
of vibration are probes of the local electronic environment. 
Vibrational frequencies can respond to molecular charge variations,
therefore detecting the 
change of ionicity ($\rho$) implied in the NIT.
Generally, $\rho$ is estimated by assuming a linear
frequency dependence upon the molecular charge,
$\omega(\rho) = \omega_0 - \rho~ \Delta$, where $\omega_0$ is
the frequency of the neutral molecule, and $\Delta$
the ionization frequency shift. A shift of at least
25 {\cm} can give reasonably accurate $\rho$ values.\cite{40years}
Vibrational spectroscopy
has been also the first investigative tool able to show that
the NIT is associated with a distortion (dimerization) of the
mixed stack.\cite{ttfcajcp}
The possibility of discriminating between regular and dimerized
stacks through vibrational spectroscopy is due to the effect
of the interaction between CT electron and the molecular
vibrations ({\it e-mv} coupling).\cite{bibbia}
We summarize here the simple symmetry arguments which are
the basis for this discrimination.

If the frontier molecular orbitals are non-degenerate, the only
molecular vibrations which can couple to the electronic excitations
are the ones belonging to the totallysymmetric ({\it ts})
representation of the molecular symmetry group.
Therefore, for a regular mixed stack
made up of centrosymmetric molecules,
no vibronic effects due to {\it e-mv} coupling can be observed in
the IR spectra,
since the molecules lie on inversion centers along the stack and the 
{\it ts} vibrations are forbidden. On the other
hand, {\it ts} modes are Raman active, so that perturbation 
effects due to the {\it e-mv} interaction are observed in Raman,
with a frequency shift of {\it ts} bands towards lower wavenumbers.
In dimerized mixed stack systems the
inversion center is lost, and
the {\it ts} modes coupled to CT are 
both Raman and IR active. In IR, they borrow
intensity from the nearby CT electronic transition, and are
therefore polarized along the stack, with an intensity
proportional to the extent of stack dimerization.\cite{bibbia}

In the case of {\CD} the situation is more complicated,
and we deal separately with the A and D vibrations.
If one disregards as reasonable approximation the slight deviation
from planarity of the ring with its substituents in {\D}, and considers
the methyl groups as point masses, then the centrosymmetric
$C_{2h}$ group can be used to classify the normal modes.\cite{lunardi}
The distribution of the vibrational modes is as follows:
\[
\Gamma_{\mathrm{{\D}}}=15a_g+6b_g+7a_u+14b_u
\]
\noindent
In the isolated molecule, the ungerade modes are IR active,
the $b_u$ being polarized 
in the molecular plane, and the $a_u$ out-of-plane, whereas the gerade modes
are Raman active. To the above modes, we have to add the 18
vibrations of the methyl group, which can be both IR and Raman active,
and polarized both in the molecular plane and normal to it.  
Since at this level of approximation {\D} is centrosymmetric, the
above considerations about the effect of {\it e-mv} coupling apply.
We note, however, that the room temperature {\CD} space group
is non-centrosymmetric
(actually, it has no symmetry elements), so that in the crystals
{\D} molecules do not reside on inversion center, and there is
no actual symmetry restriction on the IR activity of the modes.

Adopting the same criteria as in in the case of {\D}, we assume
$C_s$ molecular symmetry for {\C}, so that the distribution
of the vibrational modes is:
\[
\Gamma_{\mathrm{{\C}}}=29a'+13a''
\]
\noindent
and all modes are Raman and IR active. Totallysymmetric $a'$ vibrational
modes are polarized in the 
molecular plane (perpendicular to the stack direction), and
their frequency is affected by {\it e-mv} coupling both in regular 
and dimerized mixed stack.\cite{bibbia}
On the other hand strong vibronically activated $a'$ IR 
absorptions polarized along the stack are expected only when
the stack is dimerized. Again, the 9 vibrations due to the methyl
group can be both Raman and IR active, with unpredictable polarization.

In view of the above, the {\CD} degree of ionicity $\rho$ can
only be estimated on the basis of the {\D} $b_u$ modes,
since all the other in-plane modes are perturbed by
{\it e-mv} coupling, and the out-of-plane ones generally do not
display a linear frequency shift with ionization.\cite{40years}
The onset of dimerization is marked by the appearance of
IR bands polarized along the stack, at the same frequency of the
Raman counterpart.
The analysis of the spectral region 1350-1500 {\cm} is made
difficult due to presence of the many methyl group bending vibrations,
whose intensity may also change with temperature due to rearrangements
in the methyl groups orientation.

\subsection{Room Temperature Vibrational Spectra}

\begin{figure}[ht]
\includegraphics* [scale=0.4]{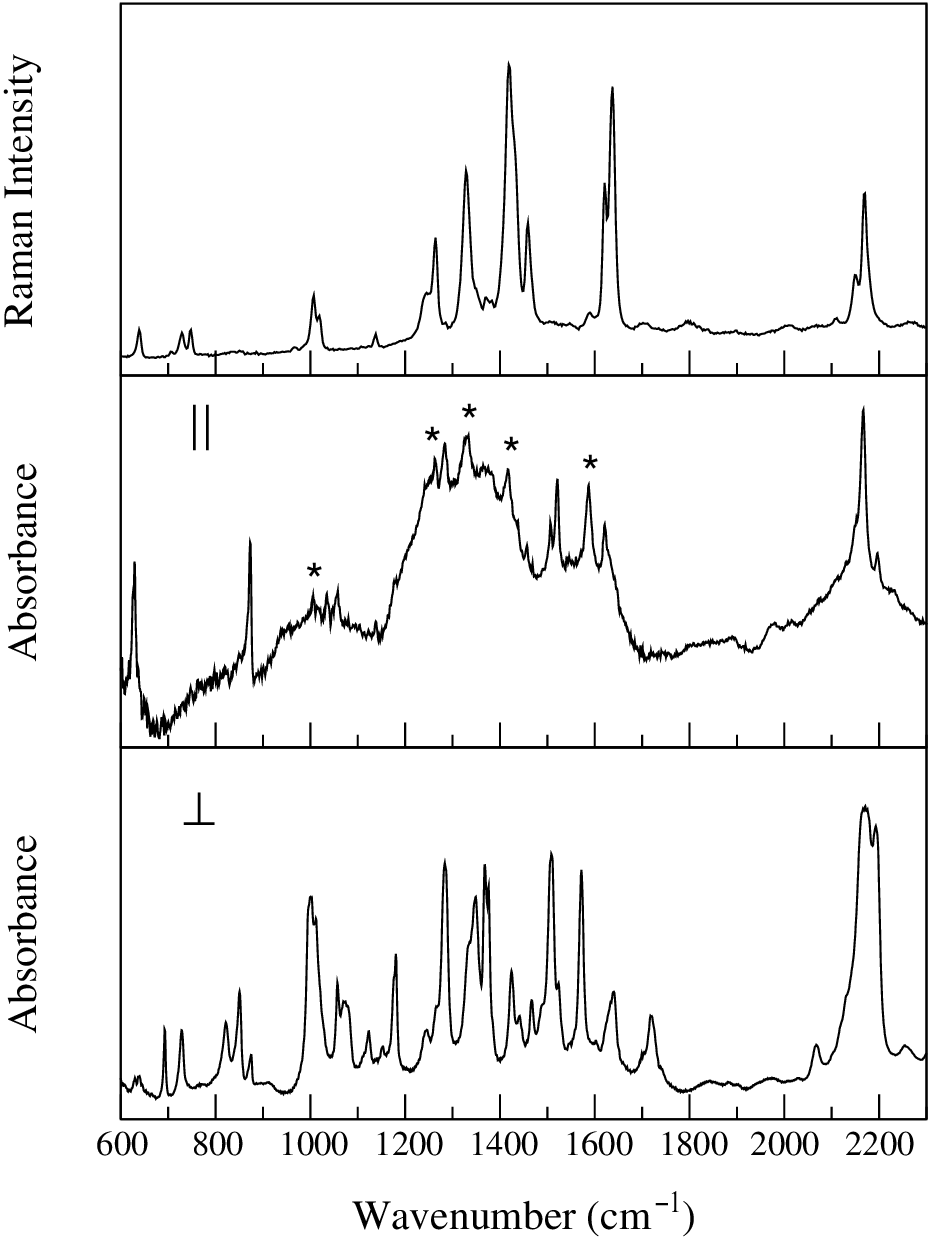}
\caption{Room temperature Raman and polarized IR spectra of
{\CD}. The IR spectra polarized parallel and perpendicular
to the stack axis are labeled with the symbols $\parallel$
and $\perp$, respectively. The {\it e-mv} induced band in the
$\parallel$ IR spectrum are marked by an asterisk (see text)}
\label{rtspectra}
\end{figure}

The room temperature Raman and polarized IR absorption
spectra of {\CD} are shown in Fig. \ref{rtspectra}.
For the sake of clarity, the polarized IR
spectra in Fig. \ref{rtspectra} refer to two different
crystals, the one relevant to the polarization
perpendicular to the stack ($\perp$, bottom panel)
being obtained on a thicker crystal. The corresponding
parallel ($\parallel$) spectrum in fact saturates completely
in the 1200-1400 {\cm}. The IR frequencies, relative intensities
and polarizations reported in Table 1 obviously refer to
the thinner crystal. The micro-Raman data do not give
polarization information. Since we shall not be
concerned with the CH and NH stretching modes, data above 2300 {\cm}
are not reported.
The interpretation of the spectra takes advantage
of the available assignments of neutral 
and fully ionized {\D} molecules.\cite{lunardi}
For {\C} a detailed vibrational assignment is not
available, and the interpretation is based on the
comparison with the frequencies of neutral {\C}
and with the normal mode analysis of  
{\it p}-phenylenediamine.\cite{hester}
The resulting assignment of {\CD} spectra is reported
in the rightmost column of Table 1.

Since the Raman exciting line is in pre-resonance with
an intramolecular electronic excitation of {\D},\cite{aoki}
we expect (and find) that the Raman bands are for the most
part due to {\D} $a_g$ modes. We remark again that the frequencies
of these modes are perturbed (lowered) by {\it e-mv}
interaction, so that they cannot be used to estimate $\rho$.

In the IR spectrum polarized parallel to the
stack direction (Fig. 1, middle panel) we notice
a couple of strong bands below 900 {\cm}, attributable
to out-of-plane modes of {\D} and {\C}. In the
spectral region 1200-1600 {\cm} one sees the clustering
of many bands, also due to the presence of the methyl
group vibrations. Despite the difficulty in interpreting this
complex spectral region, one can notice several peaks
(the main ones are marked by an asterisk in Fig. 1)
whose frequency coincides with the Raman {\it ts} bands.
As we shall see below, the intensity of these peaks increases
by lowering temperature. Therefore these bands are due to
the activation in IR of the {\it ts} modes coupled to the CT
electrons, indicating that at room temperature the stack is already
dimerized to some extent. Also the  IR band at 2167 {\cm}
has a frequency coincident with that of the Raman spectrum, and
migth be interpreted as an activated mode.
However, its strong intensity casts some doubt on this intepretation.
On the other hand, this spectral region appears rather complex
also in the spectra of neutral {\D} and of its salts.\cite{lunardi,yamakita}
In our case, in addition, the structure is disordered, and the
C$\equiv$N group is close either to the {\C} Cl or methyl group, making
plausible the presence of multiplets where a single band would be expected.

The degree of ionicity $\rho$ can be evaluated from the {\D}
$b_u$ bands appearing in the perpendicular IR spectrum.
Three modes, namely the $b_u \nu_{45}, \nu_{46}$ and $\nu_{47}$,
exhibit a ionization frequency shift larger than 25 {\cm},
and can in principle yield a good $\rho$ estimate. 
Actually, the the $b_u \nu_{45}$ (C$\equiv$N stretching)
has to be excluded from the list, since, apart from
the above mentioned difficulties\linebreak
\begin{figure}[ht]
\includegraphics* [scale=0.4]{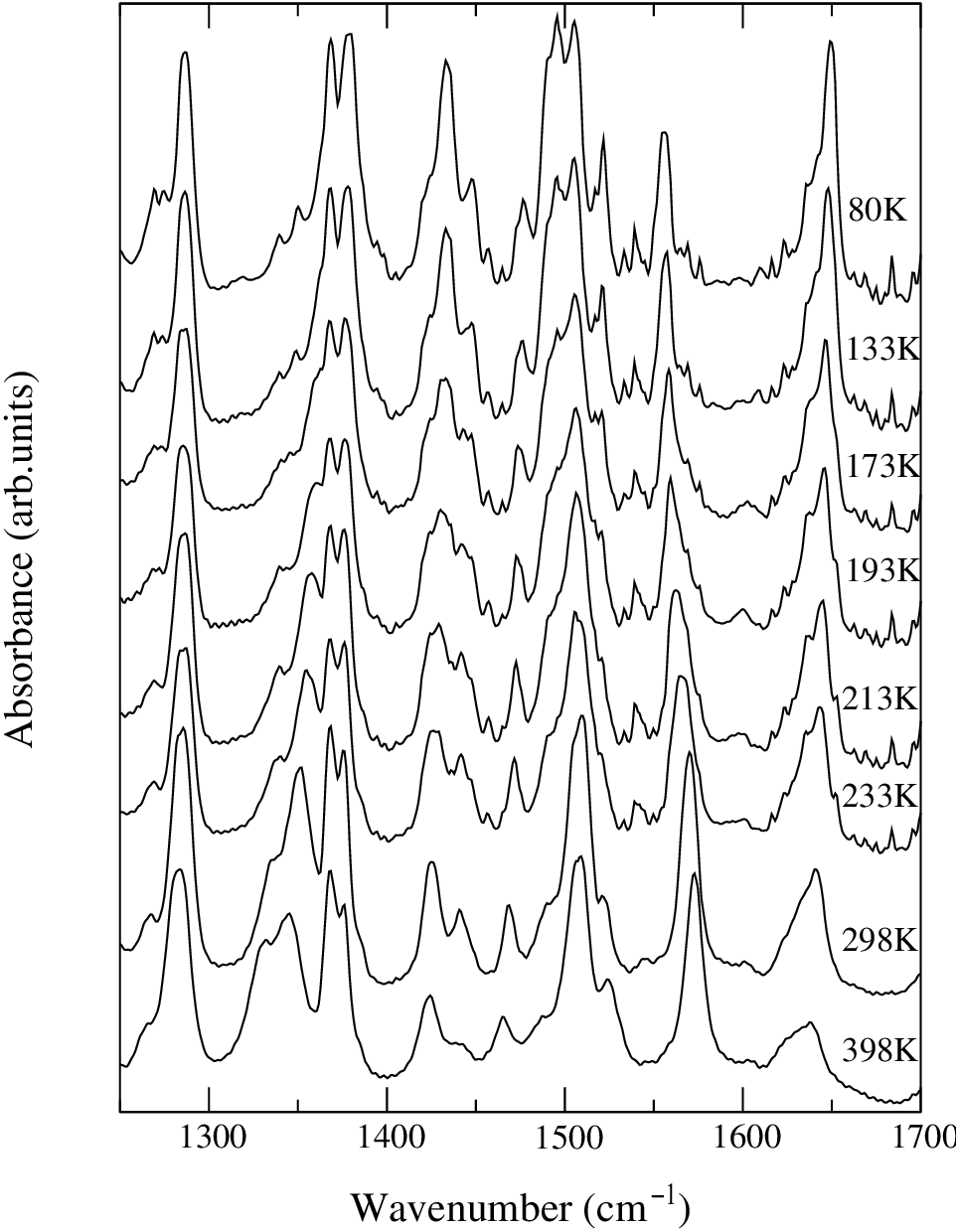}
\caption{Temperature dependence of the IR absorption spectra
polarized perpendicular to the stack. Only the spectral region
1250-1700 {\cm} is reported.}
\label{irperp}
\end{figure}

\noindent
in the interpretation
of the corresponding spectral region, its frequency is 
very sensitive to environment effects, and for instance
changes appreciably just by changing the counterion in
the {\D} alkaline salts.\cite{lunardi}
According to Lunardi and Pecile~\cite{lunardi}, the
$b_u \nu_{47}$ mode should be preferred to the $\nu_{46}$
since it is not subject to Duschinsky effects and its
frequency has a nice linear dependence on $\rho$.
On the other hand, as already pointed out by Ref. \onlinecite{aoki},
in the present case we have the interference of a very strong
$a'$ band of {\C} at 1504 {\cm} (see Fig. \ref{rtspectra}, bottom panel),
so that it is difficult to assess the frequency of the overlapping
$b_u \nu_{47}$. We exclude the assignment to the band at 1522 {\cm},
since this band is rather weak, whereas the $b_u \nu_{47}$ has
generally remarkable intensity.\cite{lunardi,kobayashi}
The $b_u \nu_{46}$ mode, on the other hand, occurs in a spectral region free
from other vibrational modes, and has a reasonably good linear dependence
on $\rho$.~\cite{kobayashi} We therefore use this mode as a primary
reference for the $\rho$ estimate, and check for compatibility
with the spectral region where the $\nu_{47}$ should occur.
On this basis we conclude that at room temperature the degree
of ionicity of {\CD} is between 0.35 and 0.44 (see below
for a more detailed discussion). We do not discuss
in detail the interpretation of the remaining parts of the
IR perpendicular spectrum, whose assignment is reported in
the rightmost column of Table 1.

\subsection{Temperature Evolution of the Spectra}

\begin{figure}[ht]
\includegraphics* [scale=0.5]{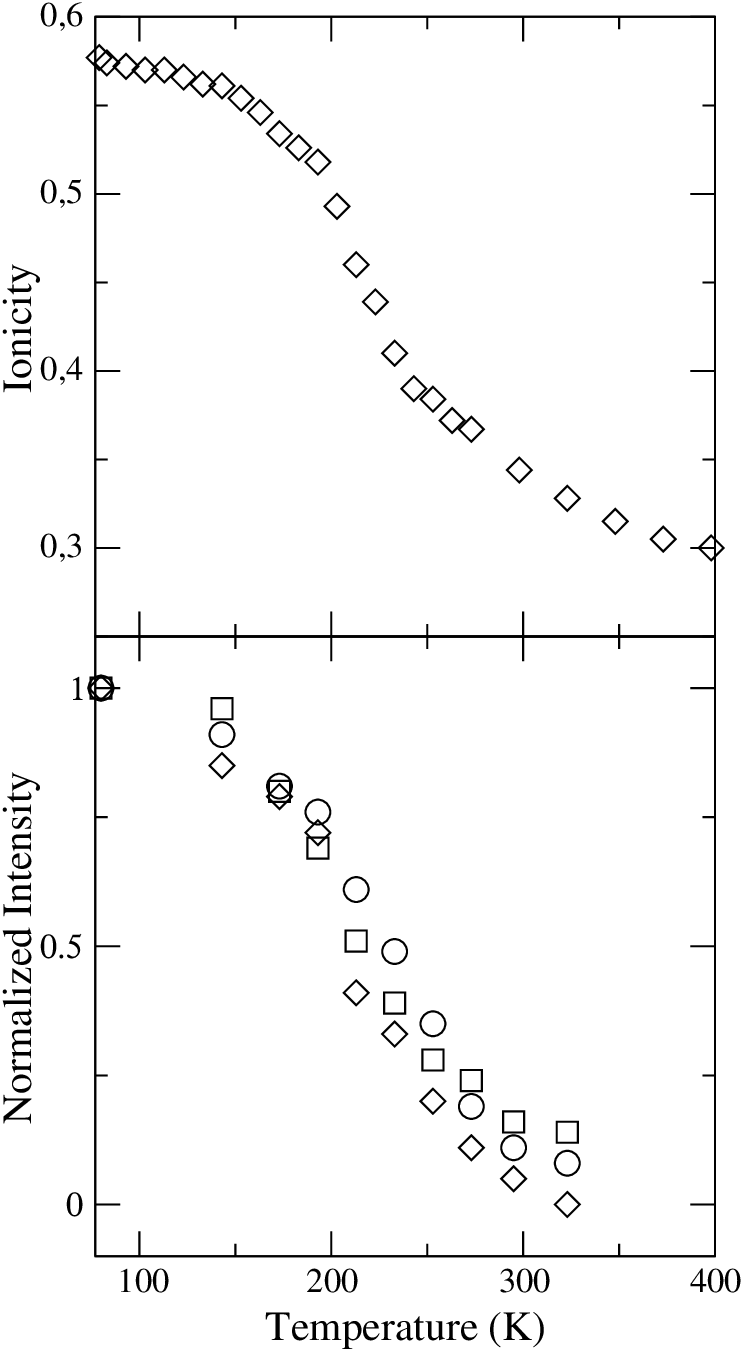}

\protect\vskip0.3truecm
\caption{Top panel: Temperature evolution of the degree of ionicity,
$\rho$, estimated from the frequency of the {\D} $b_u \nu_{46}$ mode.
Bottom panel: the corresponding change in the intensity of the {\it e-mv}
induced bands, normalized to the low-temperature value. The square,
circle and diamond correspond to the {\D} $a_g~\nu_5,~\nu_7$,
and $\nu_{13}$ modes, respectively.}
\label{rhodim}
\end{figure}

We first discuss the change in ionicity with temperature, and for this
purpose we report in Fig. \ref{irperp} the perpendicular IR spectrum
at different temperatures in the spectral region 1250-1700 {\cm}.
We immediately notice that the nearly isolated band at 1570 {\cm}, due to
{\D} $b_u$ $\nu_{46}$ mode, shows a remarkable frequency shift
towards lower wavenumbers, opposite to the usual thermal hardening 
presented by the other vibrational bands.
If we use this frequency to estimate $\rho$, as made in Ref. \onlinecite{aoki}
and discussed in the previous Section, the red shift indicates a change
in ionicity from $\sim$ 0.34 at 298 K to $\sim$ 0.57 at 80 K.
We have also increased the temperature up to 400 K (sample sublimation
starts already above 320 K), and $\rho \sim$
0.30 at this temperature. Since the spectral region of the $b_u \nu_{46}$
mode is free from overlapping with other modes, the temperature evolution
of the corresponding frequency allows us to follow the change in $\rho$,
as reported in the top panel of Fig. \ref{rhodim}. The Figure
clearly shows that $\rho$ changes smoothly with temperature.
We have not observed hysteresis. Furthermore, the simultaneous presence of
neutral and ionic domains can be excluded, since we always observe a single
band throughout all the temperature range.

We now look for the other $\rho$ diagnostic mode, {\D} $b_u \nu_{47}$,
whose presence is obscured by the strong band due to {\C} $a' \nu_{10}$
mode, located at 1508 {\cm} in the neutral molecule.
At room temperature one sees \linebreak 

\noindent
a doublet in this spectral
region, at 1504 and 1511 {\cm}, that we tentatively assign
to the {\C} $a' \nu_{10}$ and {\D} $b_u \nu_{47}$, respectively.
By lowering temperature, the doublet coalesces in a single broad band,
from which a structure grows on the low frequency side.
We consider a small hardening of the 1504 {\cm} band to
1506 {\cm} at 80 K, and assign the low frequency band (1496 {\cm}
at 80 K) to {\D} $b_u \nu_{47}$ mode. Using these wavenumbers
for the $b_u \nu_{47}$, we obtain $\rho \sim$ 0.44 at 298 K
and $\sim$ 0.69 at 80 K, compared to $\rho \sim$ 0.34
and $\sim$ 0.57 as estimated above from the $b_u \nu_{46}$ mode.
Although the absolute ionicity differ by about 0.1, the ionicity
change from 298 to 80 K is very similar, $\sim$ 0.25. 
If we take the average between the two values, we obtain 
$\rho$ = 0.39 at 298 K and $\rho$ = 0.63  at 80 K. Therefore, the
$\rho$ scale of Fig. \ref{rhodim} might be off by as much as 0.05
towards higher values. 
 
A last observation on the perpendicular spectra of Fig. \ref{irperp}
is the considerable hardening of the band at 1351 {\cm},
assigned to the {\D} $a'\nu_{12}$,
which below 170 K melts into the nearby intense doublet, 
with an estimated frequency shift of 15-20 {\cm}. 
The hardening is probably associated with the ionicity variation,
in this case towards higher frequencies, but in any case we do not use this
mode to estimate the ionicity since, as discussed in Section 3.1,
the {\C} $a'$ are perturbed by {\it e-mv} interaction.

\begin{figure}[ht]
\includegraphics* [scale=0.38]{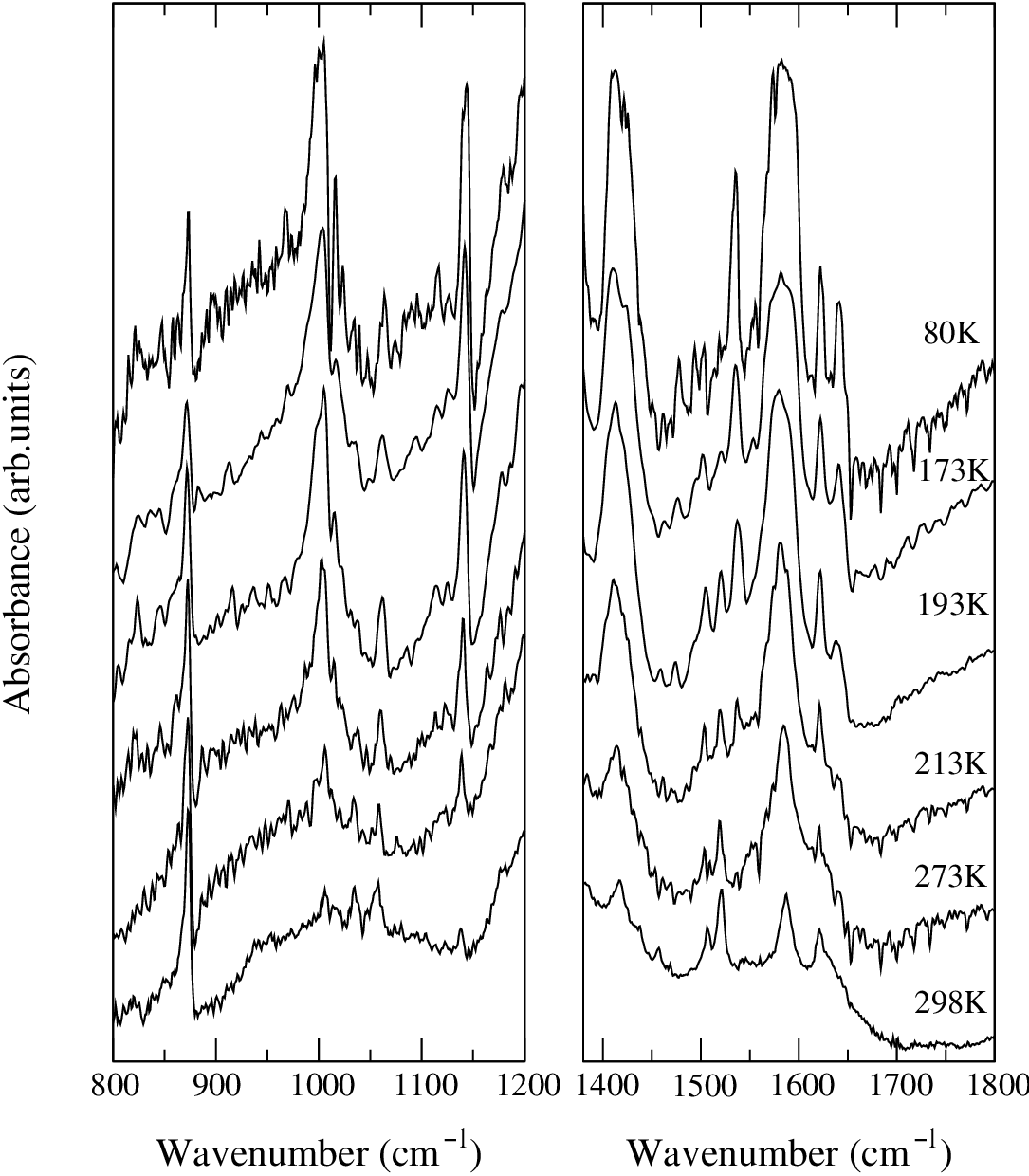}
\caption{Temperature evolution of the IR absorption spectra
polarized parallel to the stack.}
\label{irpar}
\end{figure}

Upon dimerization of the stack, the {\it e-mv} coupling yields
the intensity enhancement of the D and A {\it ts} modes in the
IR spectra polarized parallel to the stack axis. In Fig. \ref{irpar}
we report the temperature evolution of the parallel IR absorption
spectra in the 800-1200 and 1380-1800 {\cm} (the absorbance of
the frequency region in between saturates completely by lowering
temperature). The {\it e-mv} induced (vibronic) bands due
to {\it ts} modes are immediately recognized since
they gain IR intensity by lowering the temperature.
They also have their counterpart at the same frequency in the
Raman spectra (see Table 1).
The detailed evolution of the intensity enhancement of the vibronic bands
is better followed through the {\D} $a_g \nu_{13}$ mode (1006 {\cm} at room
temperature), since its intensity is easily compared with that of
the nearby band at 873 {\cm} assigned to an {\it ungerade} out
of plane ($a''$) IR  active vibration of {\C}, which does
not show any spectral variation with temperature.
In Fig. \ref{rhodim}, bottom panel, we have reported the intensity 
(normalized to the 80 K value) of this mode, together with
that of the bands
at 1417 and 1587 {\cm}, assigned
to the {\D} $a_g \nu_7$ and $\nu_5$ modes, respectively.
From the bottom panel of Fig. \ref{rhodim} we again notice
that at room temperature the {\it e-mv} bands are already
developed. This residual residual IR intensity
(about 10\% of the low temperature value) 
of the {\it ts} modes puts in evidence that there is some degree\linebreak

\noindent
of stack distortion at 298K, that is, the structural phase transition 
has already started. We have also verified that the residual 
IR intensity decreases further by increasing the temperature up to
$\sim$320 K, but we could not go further since sample sublimation
prevents the collection of reliable relative intensity data.

From Fig. \ref{irpar} we  also observe the onset of two weaker bands 
in the low temperature parallel IR spectra.
The first feature, which is still present in the room temperature
spectrum as a weak band at 1138 {\cm}, can be assigned to the weakly coupled
$a_g \,\nu_{11}$ {\it ts} mode, since exhibits a frequency coincident    
Raman counterpart.
The second spectral feature appears in the low temperature spectrum as a
doublet around 1535 {\cm}, in a region where no fundamental modes 
are expected. Therefore, overtones and combinations 
are likely involved in the doublet observed in this region.

\begin{figure}[ht]
\includegraphics* [scale=0.45]{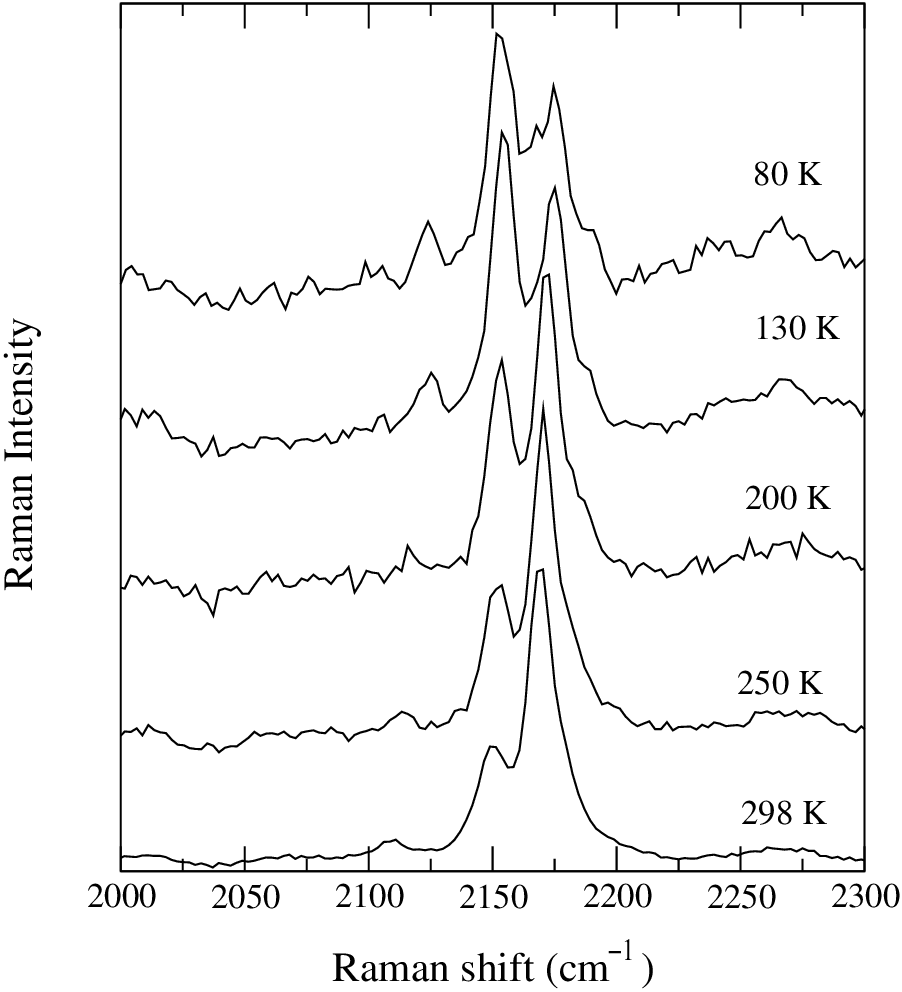}
\caption{Temperature evolution of the Raman spectra
in the C$\equiv$N stretching frequency region.}
\label{raman}
\end{figure}

The C$\equiv$N stretching region deserves a separate discussion.
In Fig. \ref{raman} we present the temperature evolution of the Raman
spectra in the corresponding frequency range.
We have already remarked that at room temperature Raman and IR
spectra present more bands than expected.
As shown in Fig. \ref{raman}, the room temperature Raman spectra
shows two bands. By lowering temperature, 
the high frequency band at 2168 {\cm} decreases
in intensity, but remains present also at 80 K, hardening to 2175 {\cm},
whereas the low frequency band grows. 
Moreover, we observe the onset of a third band at still lower frequency.
We also find a counterpart
of the above three low temperature Raman bands in the parallel
IR spectrum, indicating that they correspond to {\it e-mv}
coupled modes. At present, we are unable to interpret this spectral
region, where, as mentioned above, disorder effects may be important.
\vskip0.5truecm

We finally remark that several other minor features and intensity
changes appearing below about 200 K both in IR and Raman
spectra are not interpreted. They
might be due to some molecular rearrangments following the
ionicity change, but at the moment this remains a speculative
hypothesis.

\section{Discussion}

As we have seen in the previous Section, the {\CD} vibrational
spectra are rather complex, and their assignment is not obvious.
The parallel use of Raman and polarized IR
spectra as function of temperature helps us to reach a satisfactory
spectral interpretation. Our data essentially confirm the previuosly
given analysis of {\CD} TINIT,~\cite{aoki} but the detailed study
presented here offers a more extensive picture and puts
in evidence new and interesting
features of the N-I interface in this compound.

As it is immediately evident from the top panel of
Fig. \ref{rhodim}, the {\CD} ionicity changes continuously with
lowering $T$. 
The ionicity change from high to low temperature
turns out to be about 0.3, of the same order of magnitude as
in TTF-CA.~\cite{ttfcajcp} In TTF-CA, on the other hand, the
temperature evolution of $\rho$ is different: by lowering
$T$ at ambient pressure, $\rho$ increases continuously
from about 0.22 to about 0.30, and at $T_c$ = 82 K there is a first
order transition to $\rho \sim $ 0.64. The first order
nature of the phase transition is confirmed by several evidences,
among which we cite the small hysteresis,\cite{nonso} and the
sometimes observed simultaneous presence of N and I domains
in proximity of the phase transition. The latter fact is
evidenced by the presence of IR and Raman bands due to both
neutral and ionic species.\cite{ttfcajcp} On the opposite,
in {\CD} the passage from neutral to ionic side is smooth,
and we do not find spectral signatures of
different domains.

We remark that {\it all} the so far known TINIT's exhibit
a discountinuos $\rho$ jump at $T_c$, albeit in some cases
the jump is rather small.\cite{iwasa1990} Another feature common
to all ambient pressure TINIT's is sudden dimerization of the
stack at $T_c$. Also in this respect the behavior of {\CD}
is remarkably different. As shown in Fig. 1 and discussed
in the previous Section, the {\CD} stack is already dimerized
to some extent at room temperature. The intensity of the
{\it e-mv} induced IR bands, which is related to stack
distortion, has the same S-shape as the $\rho$ variation
(see Fig. \ref{rhodim}). The point of maximum slope occurs in
both cases around 220 K.
Given the already mentioned uncertainty in the
absolute $\rho$ value, we cannot say whether or not
the crossing from N to I side also occurs around
220 K. On the other hand, $\rho$ can be considered an order
parameter only in the case of first order, discontinuous
phase transitions. Furthermore, if $\rho$ changes
continuously, the N-I borderline is ill-defined:
From a theoretical point of view, the borderline
is at $\rho$=0.5 in the limit of an isolated DA pair,
and at $\rho$ = 0.64 for a regular stack.\cite{VB}
In any case, it is clear that for {\CD} the dimerization
phase transition is already started when the compound is
still well into the N phase. 
  
The commonly reported picture of TINIT is as follows.
The crystal contraction induced by lowering $T$ increases
the crystal's Madelung energy, up to the point of inducing a first
order NI phase transition. On the ionic side, the system is
considered to be intrinsically unstable towards dimerization
due to spin-Peierls mechanism, so that the NI transition is
accompanied by stack dimerization. Although the above picture
explains the salient features of the so far known TINIT's,
it represents a simplified view, since it considers only
the small CT integral ($t$) limit, and disregards the role
of electron-lattice phonon {\it e-lph} coupling.
Certainly, the {\CD} phase transition cannot be explained in the
above terms. Zero-temperature phase diagrams based on valence
bond exact diagonalization methods, corroborated
by experimental observations on several mixed stack CT
complexes, provide a more complete and detailed picture,\cite{VB}
and allow us to rationalize the {\CD} phase transition.

First of all, the dimerization transition is
attributable to the spin-Peierls mechanism only in the case of 
fully ionic ($\rho \cong $ 1) CT complexes like
TTF-BA,~\cite{ttfba} namely in the large $\delta/t$
limit ($\delta$ is the difference between D and A site
energies).\cite{VB} By decreasing the ionicity, the\linebreak

\vskip0.8truecm

\begin{figure}[ht]
\includegraphics* [scale=0.5]{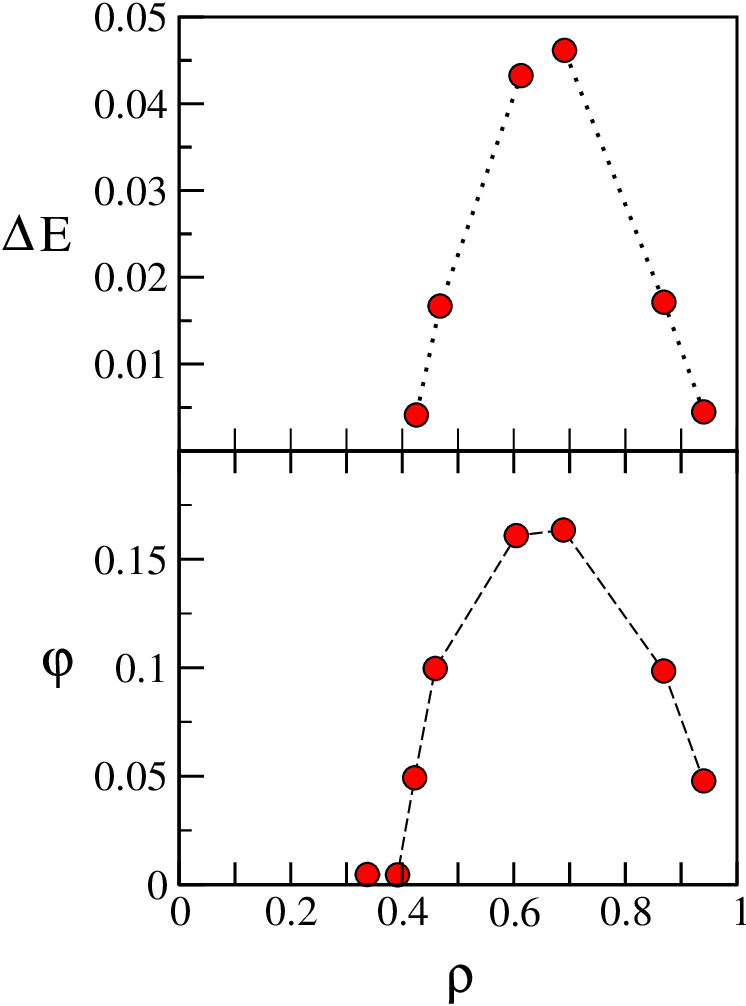}
\caption{Top panel: calculated energy gain upon dimerization
for a mixed stack as function of $\rho$. Bottom panel:
the corresponding variation of the asymmetry parameter
(from Ref. \protect\onlinecite{VB}).}
\label{VBFig}
\end{figure}

\noindent
charge modulation acquires importance, and the dimerization
is better classified as an ordinary Peierls transition.
The top panel of Fig. \ref{VBFig} shows the calculated energy gain
upon dimerization.\cite{VB} The gain is maximum
for intermediate ionicity, corresponding to a pure Peierls
mechanism, and progressively decreases as $\rho \rightarrow$ 1,
corresponding to pure spin-Peierls. 
We can then understand why the 
temperature of the dimerization transition increases with
decreasing $\rho$ from 1 to intermediate values: it is 220 K
for TMPD-TCNQ ($\rho \sim $
0.9),\cite{tmpdtcnq} and TMPD-CA ($\rho \sim $ 0.6)
is already dimerized at room temperature.\cite{tmpdca}
Finally, if the {\it e-lph} coupling is strong enough, the
dimerization transition may occur also on the neutral side, as
in the case of DBTTF-TCNQ under pressure.\cite{dbttftcnq}
Whereas the dimerization transition is induced by {\it e-lph}
coupling, the $\rho$ variation is driven by the Madelung energy,
and is affected by the interplay of $\delta$, the electron-
molecular vibration {\it e-mv} coupling, and the CT integral.
Since the $\rho$ variation does not imply a symmetry change,
we can assimilate the NI interface to the gas-liquid one,
where we have a critical point beyond which there is no
phase boundary.\cite{VB}

We then propose the following scenario for the {\CD}
phase transition. As pointed out in Ref. \onlinecite{aoki},
the CT integral in this compund is higher than in TTF-CA.
Since the position of the critical point for the passage
from a discontinuous to a continuous NI interface depends
on the ratio of the Madelung potential to the CT integral,
we might be beyond the critical point, so that the
interface is continuous and $\rho$
is not a proper order parameter. In any case, at room temperature
$\rho \geq$ 0.35 (in TTF-CA $\rho \cong $ 0.3 when the NIT occurs),
and for such high $\rho$ the gain in energy upon dimerization~\cite{VB}
is such that the dimerization transition already starts around 320 K.
The dimerization phase transition is very broad, about 200 K,
and is in many aspects similar to that of TMPD-TCNQ.~\cite{tmpdtcnq}
TMPD-TCNQ, on the other hand, is already ionic, and the $\rho$
change during the dimerization transition, if any, is very
small.\cite{tmpdtcnq} For {\CD} the dimerization transition
starts on the neutral side, and in such a case we expect
that the extent of dimerization increases with $\rho$.
Fig. \ref{VBFig}, bottom panel, shows the $\rho$
dependence of the asymmetry parameter,
$\phi \propto (t_i - t_{i+1})/(t_i + t_{i+1})$, as obtained
by VB calculation for a set of parameters relevant to systems
exhibiting continuous ionicity change.\cite{VB} It is seen
that $\phi$, which is clearly related to the extent of dimerization,
increases as the ionicity increase, reaching the maximum for $\rho$ between
0.6 and 0.7. Thus the dimerization represents the order parameter
of the phase transition, but it is driven by the degree of ionicity
(assuming constant the strength of {\it e-lph} coupling).

The above picture is highly reminiscent of what may occur
in Langmuir monolayers, where one has the coupling between a first-order
phase transition (order parameter: the area per molecule) and
a second order, orientational phase transition.\cite{zhong}
An analysis in terms of Ginzburg-Landau theory of coupled phase
transitions is beyond the aim of the present paper, but represents
one of the most interesting developments in the study of {\CD}
phase transition. Further experiments, including high pressure
measurements, are in progress to confirm and extend the present
results.



\acknowledgements

We acknowledge
helpful discussions with A.Painelli.
This work has been supported by the Italian National
Research Council (CNR) within its ``Progetto Finalizzato
Materiali Speciali per tecnologie Avanzate II'', and the
Ministry of University and of Scientific and Technological
Research (MURST).




\eject
\onecolumn
\newpage

\scriptsize
\begin{table}
\caption{Infrared and Raman spectra of ClMePD-DMeDCNQI from 200 to 2300 cm$^{-1}$.}
\vskip 3mm
\begin{tabular}{cccccc}
\multicolumn{3}{c}{Infrared} & \multicolumn{2}{c}{Raman} &
\multicolumn{1}{c}{Assignment} \\
\cline{1-3}\cline{4-5} \\

$\tilde{\nu}$/cm$^{-1}$ (298 K) & $\tilde{\nu}$/cm$^{-1}$ (80 K)  &Polariz.&~~
$\tilde{\nu}$/cm$^{-1}$ (298 K)& $\tilde{\nu}$/cm$^{-1}$ (80 K)& \\ \\
\hline \\
     &      &         & 279w & 279 & $A, \, a_g \nu_{19}/\ D, \, a'\nu_{28} $\\ 
     &      &         & 361m & 363 & $A, \, a_g \nu_{18}/\ D, \, a'\nu_{26} $\\  
589m & 595w & $\perp$ &  &  & $A,\, b_u \nu_{57}$\\
625sh &  & $\parallel$ &  &  & \\
629s & 628m & $\parallel$ &  & 631w & $D, \, a''$\\ 
     & 644ms & $\parallel$ & 637w & 644w & $A, \, a_g \nu_{15}$\\ 
693w & 694w & $\perp$ &  &  & $A, \, b_u \nu_{56}$\\ 
729w & 729w & $\perp$ & 729w &      & \\
     &      &         & 746w & 754w & $D, \, a'~\nu_{22}$\\  
824w & 828w & $\perp$ &  &  & $D$ ?\\ 
852m & 858m & $\perp$ &  &  & $A,\, b_u \nu_{55}$\\ 
873s & 873m & $\parallel,\,\perp$ &  &  & $D, \, a''$\\ 
996m & 997m & $\perp$ &  &  & $D, a'~\nu_{20}$\\
1002m & 1005w & $\perp$ &  &  & $A, \, b_u \nu_{54}$\\ 
1006w & 1000s  & $\parallel$ & 1006m & 1000s & $A, \, a_g \nu_{13}$\\ 
1012sh & 1017m & $\perp$ &  &  & $D, a'\nu_{19}?$\\ 
1017vw & 1016w  & $\parallel$  & 1018w & 1025w & $A, \, a_g \nu_{12}$\\ 
1035w &       & $\parallel$  &  &  & \\ 
1056w & 1064w & $\parallel$  &  &  & \\ 
1059w & 1068w & $\perp$ &  &  & $A, b_u\nu_{53}/\ D, a'\nu_{18}$\\ 
1073br & 1084vw & $\perp$ &  &  & $D, \, a'\nu_{17}$\\ 
      &       &              &       & 1122w & \\ 
1138w & 1144s & $\parallel$  & 1137w & 1144 & $A, \, a_g \nu_{11}$?\\ 
1154vw &       & $\perp$ &  &  & \\ 
1176sh & 1179m & $\perp$ &  &  & $D, \, a'\nu_{16}$\\   
1181m & 1186w & $\perp$ &  &  & $A, \, b_u \nu_{52}$\\  
      & 1230s & $\parallel$ & 1245sh & 1231vs & $A, \, a_g \nu_{10}$\\ 
1263m &       & $\parallel$  & 1263m & 1255w & \\  
1266sh & 1269sh & $\perp$ &  &  & \\  
1285s & 1286m & $\perp$, $\parallel$ &  & & $A, \, b_u \nu_{51}$\\ 
1329m & 1325vs & $\parallel$ & 1328s & 1322s & $A, \, a_g \nu_{9}$\\ 
1335sh & 1350vw & $\perp$ &  &  & \\ 
1351m & (1368)  & $\perp$ &  &  & $D, a'\nu_{12}$\\ 
1368ms & 1368m & $\perp$ &  &  & \\ 
1376m & 1379m & $\perp$ &  &  & $A, \, b_u \nu_{50}$\\ 
1417m & 1412vs & $\parallel$ & 1418vs & 1412s & $A, \, a_g \nu_{7}$\\ 
1425m & 1434m & $\perp$ & 1428sh  &  & $D, \, a'\nu_{11}$\\  
1441w & 1448vw & $\perp$ &  &  & $A, \, b_u \nu_{48}$\\ 
1457w &       & $\parallel$ & 1458m &  & \\ 
1468w & 1477w & $\parallel$ &  &  & \\ 
1489sh & 1490sh & $\perp$ &  & 1489w  & \\ 
      & 1496s & $\perp$, $\parallel$ &  &  & $A, \, b_u \nu_{47}$?\\  
1504s & 1506s & $\perp$, $\parallel$ &  &  & $D, \, a'\nu_{10}$\\  
1507w &  & $\parallel$ &  &  & \\ 
1511s &  & $\perp$ &  &  & $A, \, b_u \nu_{47}$?\\ 
1522w &  & $\perp$, $\parallel$ &  &  & \\ 
      & 1522vw & $\perp$ &  &  & \\
      & 1532w & $\parallel$ &  & 1534w & \\
      & 1536w & $\parallel$ &  & &  \\
      & & & & 1553w &  \\ 
1570m & 1556m & $\perp$ &  &  & $A, \, b_u \nu_{46}$\\ 
1587m & 1582vs & $\parallel$ & 1586vw & 1580s & $A, \, a_g \nu_{5}$\\ 
1621br & 1622w & $\parallel$ & 1619w & 1623s & $D, \, a'$\\ 
      &  &  & 1637s & 1640s & $A, \, a_g\nu_5$\\ 
1640w & 1649m & $\perp$ &  &  & $D, a' \nu_7$\\ 
1721br &  & $\perp$ &  &  & \\
      & 2085w  & $\perp$ &  & & \\ 
      & 2124m  & $\parallel$ &  & 2124w &  $A, \, a_g \nu_{4}$ ?\\
      & 2138w  & $\perp$ &  & &  \\ 
2150sh & 2153vs & $\parallel$ & 2148w & 2153s &  $A, \, a_g \nu_{4}$\\ 
2167s & 2175s & $\parallel$ & 2168s & 2175m & $A, \, a_g \nu_{4}$\\ 
2171vs & 2171s & $\perp$ &  &  & $A, \, b_u \nu_{45}$\\ 
2197w & 2213w & $\perp$, $\parallel$ &  &  & \\ 
 
\end{tabular}
\end{table}





\references

\bibitem{torrance1981}
J.B.Torrance, J.E.Vazquez, J.J.Mayerle, and V.Y.Lee,
Phys.Rev.Lett., 1981, {\bf46}, 253; J.B.Torrance, A.Girlando,
J.J.Mayerle, J.I.Crowley, V.Y.Lee, P.Batail, and S.J.LaPlace,
Phys.Rev.Lett., 1981, {\bf47}, 1747.

\bibitem{mconnell1965}
H.M.McConnell, B.M.Hoffman, and R.M.Metzger,
Proc.Natl.Acad.Sci. USA, 1965, {\bf53}, 46.

\bibitem{VB}
A.Painelli and A.Girlando, Phys.Rev. B, 1988, {\bf37}, 5748 and references
therein.

\bibitem{koshihara1999}
Shin-ya Koshihara, Y.Takahashi, H.Sakai, Y.Tokura,
ans T.Luty, J.Phys.Chem., 1999, {\bf103}, 2592.

\bibitem{cailleau1997}
M.H.Lem\'ee-Cailleau, M.Le Cointe, H.Cailleau, T.Luty,
F.Moussa, J.Roos, D.Brinkmann, B.Toudic, C.Ayache, and N.Karl,
Phys.Rev.Lett., 1997, {\bf79}, 1690.

\bibitem{aoki93}
S.Aoki, T.Nakayama, and A.Miura, Phys. Rev. B, 1993, {\bf48}, 626.

\bibitem{horiuchi1998}
S.Horiuchi, R.Kumai, and Y.Tokura, J.Am.Chem.Soc., 1998,
{\bf120}, 7379.

\bibitem{iwasa1990}
Y.Iwasa, T.Koda, Y.Tokura, A.Kobayashi, N.Iwasawa, and G.Saito,
Phys. Rev. B, 1990, {\bf42}, 2374.

\bibitem{aoki}
S.Aoki and T.Nakayama, Phys.Rev. B, 1997, {\bf56}, R2893.

\bibitem{40years}
C.Pecile, A.Painelli, and A.Girlando
Mol. Cryst. Liq. Cryst., 1989, {\bf 171}, 69.

\bibitem{ttfcajcp}
A.Girlando, F.Marzola, C.Pecile, and J.B.Torrance
J.Chem.Phys., 1983, {\bf 79}, 1075.

\bibitem{bibbia}
A.Painelli and A.Girlando,
J.Chem.Phys., 1986, {\bf 84}, 5655.

\bibitem{lunardi}
G.Lunardi and C.Pecile, J.Chem.Phys., 1991, {\bf95}, 6911.

\bibitem{hester}
E.E.Ernstbrunner, R.B.Girling, W.E.L.Grossman, E.Mayer,
K.P.J.Williams, and R.E.Hester,
J.Raman Spectroscopy, 1981, {\bf 10}, 161.

\bibitem{yamakita}
Y.Yamakita, Y.Furukawa, A.Kobayashi, M.Tasumi, R.Kato and H.Kobayashi,
J.Chem.Phys., 1994, {\bf100}, 2449.

\bibitem{kobayashi}
H.Kobayashi, A.Miyamoto, R.Kato, F.Sakai, A.Kobayashi, Y.Yamakita,
Y.Furukawa, M.Tasumi and T.Watanabe, Phys.Rev. B, 1993, {\bf 47}, 3500.

\bibitem{nonso}
M.H.Lemee-Cailleau, B.Toudic, H.Cailleau, F.Moussa, M.Le Cointe, G.Silly,
and N.Karl,
Ferroelectrics, 1982, {\bf 127}, 19.

\bibitem{ttfba}
A.Girlando, C.Pecile, and J.B.Torrance,
Solid State Comm., 1985, {\bf 54}, 753.

\bibitem{tmpdtcnq}
A.Girlando, A.Painelli, and C.Pecile,
Mol. Cryst.Liq.Cryst., 1984, {\bf 112}, 325.

\bibitem{tmpdca}
A.Girlando, A.Painelli, and C.Pecile,
J.Chem.Phys., 1988, {\bf 89}, 494.

\bibitem{dbttftcnq}
A.Girlando, C.Pecile, A.Brillante, and K.Syassen,
Synth. Metals, 1987, {\bf 19}, 503.

\bibitem{zhong}
Fan Zhong, M.Jiang, D.Y.Xing, and Jinming Dong,
J.Chem.Phys., 2000, {\bf 113}, 4465.
\end{document}